# Jahn-Teller Distortion-Driven Lattice Parameter Tunability in Coherent High-Entropy Oxide Thin Films

Dhiya Srikanth[1], Joseph Petruska[1], Matthew Furst[1], Kaylin Lamaute[1], Jon-Paul Maria[1] and *Saeed S. I. Almishal[1]

[1]*Department of Materials Science and Engineering, The Pennsylvania State University, University Park, PA 16802, USA*

**Corresponding Authors:** Saeed S. I. Almishal ssa5409@psu.edu

## Abstract

Here, we isolate how Cu concentration and Jahn-Teller distortions control the out-of-plane lattice parameter of pseudomorphic MgCoNiCuZnO-derived high-entropy oxide films on MgO. We prepare nine compositions spanning 0–20% Cu, verify target composition and phase by X-ray fluorescence and diffraction, and grow films with variable Cu content by pulsed laser deposition across multiple substrate temperatures. At 500 °C, the lattice parameter increases nearly linearly with Cu content; at 300 °C, it decreases nonlinearly, producing a 2.6% difference between temperature regimes for equimolar MgCoNiCuZnO at our growth conditions. We attribute this reversal to temperature-dependent coupling between Cu-driven Jahn-Teller distortions and Co valence change. While prior work identifies Co valence as the primary driver of the MgCoNiCuZnO lattice response with temperature; our results establish Cu Jahn-Teller distortion as an equally important structural control parameter.

## Introduction

High-entropy oxides (HEOs) are crystalline solid solutions containing multiple metal cations with a disordered distribution across symmetry-equivalent lattice sites while retaining long-range crystallographic order[1–7]. At sufficiently high temperatures, configurational entropy can stabilize the disorded crystalline solid solution state; rapid cooling then traps the high-entropy structure as a metastable macrostate[1–7]. The prototypical HEO MgCoNiCuZnO, commonly referred to as J14, adopts the rock salt structure at 950 °C[1,2]. Quenching preserves this structure as a kinetically trapped, metastable state at room temperature, although Cu, Zn, and Co oxides preferentially adopt other room-temperature structures[1–3]. Pulsed laser deposition (PLD) provides a far-from-equilibrium route to further control this metastable macrostate[2,8–15]. PLD rapidly quenches plasma-generated species at the substrate, kinetically arresting a potentially homogeneous solid solution and enabling diverse atomic and electronic configurations within single-phase J14 films[2,8–11]. Remarkably, Kotsonis and coauthors reported that increasing the substrate temperature by only 200 °C produces an approximately 3% increase in J14 lattice parameter at their growth conditions, attributed to changes in Co valence[9]. Yet, despite this pronounced lattice response and the underlying chemical and structural disorder, the films remain highly crystalline and pseudomorphic. This counterintuitive coexistence of substantial local disorder with robust long-range crystallographic order has recently been termed *anomalous crystallinity*[10].

The prevailing picture assigns Co the dominant, and largely independent, role in controlling the J14 lattice parameter with substrate temperature[2,9,10]. Lower growth temperatures oxidize a significant fraction of $Co^{2+}$ to $Co^{3+}$ [2,9,10]. Because the ionic radius of high-spin octahedral $Co^{3+}$ is approximately 18% smaller than that of high-spin octahedral $Co^{2+}$, it leads to lattice contraction[2,9,10]. Recent observations challenge this Co-centered view and motivate a new hypothesis: local, orientable strain from Jahn-Teller-distorted $CuO_6$ octahedra may act in concert with Co and its valence to produce the large lattice change in J14. Three findings point toward such coupling[16–19]. First, Cu-free MgCoNiZnO films show the least abrupt unit-cell volume increase with substrate temperature, suggesting that the local anisotropy of Jahn-Teller-distorted $CuO_6$ octahedra may sharpen the structural transition in J14[2,10]. Second, Cu appears to help J14 accommodate the severe size mismatch introduced by 10% $Ca^{+2}$ cations[12,16]. The 10%Ca Cu-containing film maintains pseudomorphic growth despite a 4.2% larger out-of-plane lattice parameter than its Cu-free counterpart, whereas the Cu-free film relaxes and loses coherency[12]. Third, slow deposition promotes Cu-rich coherent nanotweeds whose morphology reflects the anisotropic, stress-free transformation strain generated by tetragonal $CuO_6$ octahedra within the rock salt matrix[11,17]. Together, these observations prompt us to reconsider Co as the sole driver and instead explore how Co and Cu may cooperatively control lattice parameters through mechanisms that remain unresolved[10–12].

Here, we isolate how Cu content influences the out-of-plane lattice parameter of pseudomorphic $Cu_x(MgCoNiCuZn)_{1-x}O$ films. We hypothesize that in coherent epitaxial thin films, where the in-plane lattice parameter is pinned to the substrate, incrementally increasing the Cu concentration in MgCoNiZnO will intensify the Jahn-Teller effect (because Cu is a highly active Jahn-Teller ion) and manifest as an increase in the out-of-plane lattice parameter. One might naively expect this trend at any growth temperature. To test this prediction, we grow $Cu_x(MgCoNiCuZn)_{1-x}O$ films using PLD from calibrated targets spanning (x=0) to (x=0.20) Cu in increments of 0.025 at substrate temperatures from 200 to 500 °C and calculate the out-of-plane lattice parameter from X-ray diffraction. We choose this range because it spans the strong lattice constant transition temperature in J14. The films grown at 500 °C follow the expected Cu-driven expansion. Contrary to this expectation, the films grown at 300 and 200 °C do not. This breakdown of the intuitive trend reveals that Cu-induced strain cannot act independently of the temperature-dependent Co valence and instead points toward coupled Co-Cu control of the J14 lattice parameter.

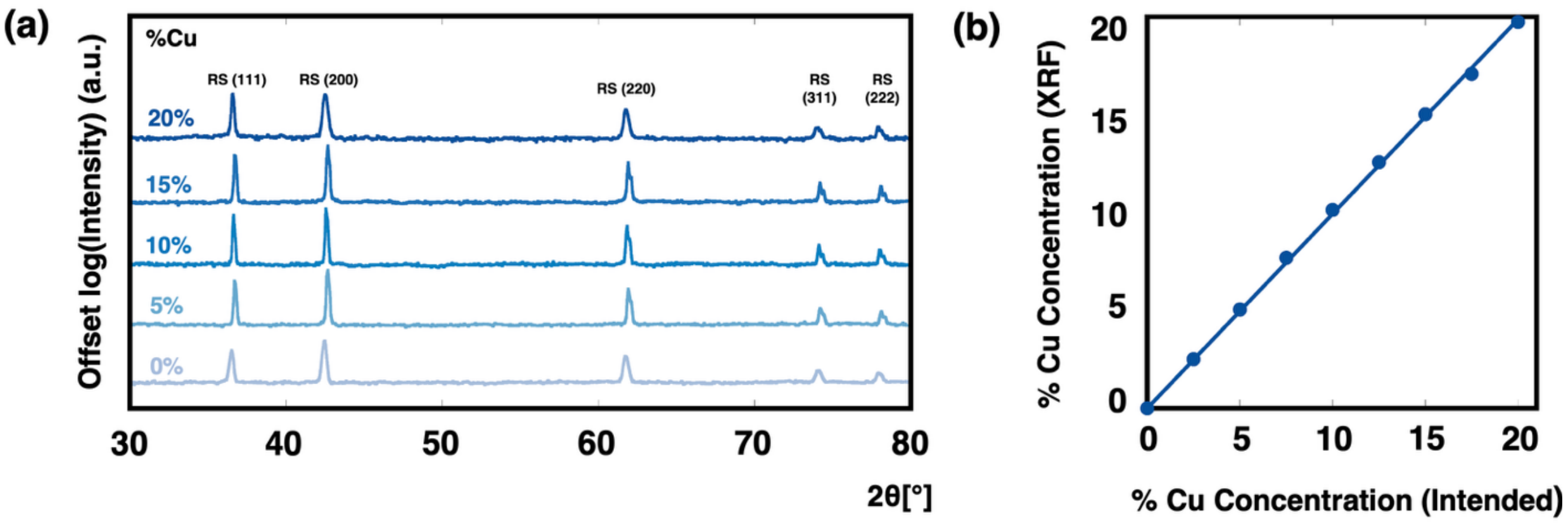

**Figure 1: Structural and compositional characterization of Cu-varying $Cu_x(MgCoNiCuZn)_{1-x}O$ bulk ceramic targets.** (a) X-ray diffraction patterns of bulk ceramic targets with nominal Cu concentrations ranging from 0% to 20%. (b) Comparison of the nominal Cu concentrations with those determined by X-ray fluorescence (XRF) fitting.

## Results and Discussion

We begin by synthesizing $Cu_x(MgCoNiZn)_{1-x}O$ bulk targets and confirming their intended compositions. Figure 1(a) shows the X-ray diffraction (XRD) patterns of the Cu-concentration series representative set, arranged from 0% Cu at the bottom to 20% Cu at the top in 5% increments. At 20% Cu, the composition becomes the prototypical five-component equimolar rock salt J14. Every bulk ceramic target adopts a single-phase rock salt structure under our synthesis conditions. We verify each target's Cu concentration using our in-house calibrated X-ray fluorescence (XRF). Figure 1(b) compares the XRF-measured Cu concentrations with the intended values. Their close agreement confirms accurate target compositions and compositional control in 2.5% Cu increments.

We then use these ceramic targets and far-from-equilibrium PLD to grow coherent epitaxial thin films on MgO and test Cu-driven Jahn-Teller lattice expansion. We grow each film using a laser fluence of 2 J/cm² and a 10 Hz pulse rate, then rapidly quench it to metastabilize the high-symmetry, high-entropy phase.[2,8–11] We limit the film thickness to approximately 80 nm to suppress the local nanostructure evolution reported previously.[11,17] We first examine the series grown at 500 °C. Figure 2(a) shows θ-2θ Bragg-Brentano high-definition X-ray diffraction (BBHD XRD) scans for this Cu-concentration series. The patterns progress from 0% Cu at the bottom to 20% Cu, or equimolar J14, at the top in 2.5% increments. Only the (002) and (004) film reflections appear across the series, confirming highly oriented, single-phase rock salt films. Both reflections shift toward lower 2θ as the Cu concentration increases, revealing the predicted increase in the out-of-plane lattice parameter. Figure 2(b) highlights this shift through high-resolution XRD scans of the film and MgO (002) peaks. The film (002) peak shifts from 2θ = 42.621° for the 0% Cu film, corresponding to an out-of-plane lattice parameter of 4.239 Å, to 2θ = 42.241° for the 20% Cu film, corresponding to 4.276 Å. We attribute this increase in out-of-plane lattice parameter to the increasing concentration of Jahn-Teller-active Cu and the resulting local distortions. At 500 °C, we expect Co to remain predominantly $Co^{2+}$, so Co valence changes do not likely drive this Cu-dependent expansion.[9,10] It is important to note that all films are pseudomorphic to the MgO substrate, thus unit cell volume is also increasing with %Cu. Finally, as shown in Figure 2(b) persistent Pendellösung fringes are present for every film, indicating uniform thickness and smooth interfaces despite the varying Cu content - further evidence of anomalous crystallinity across the series[10].

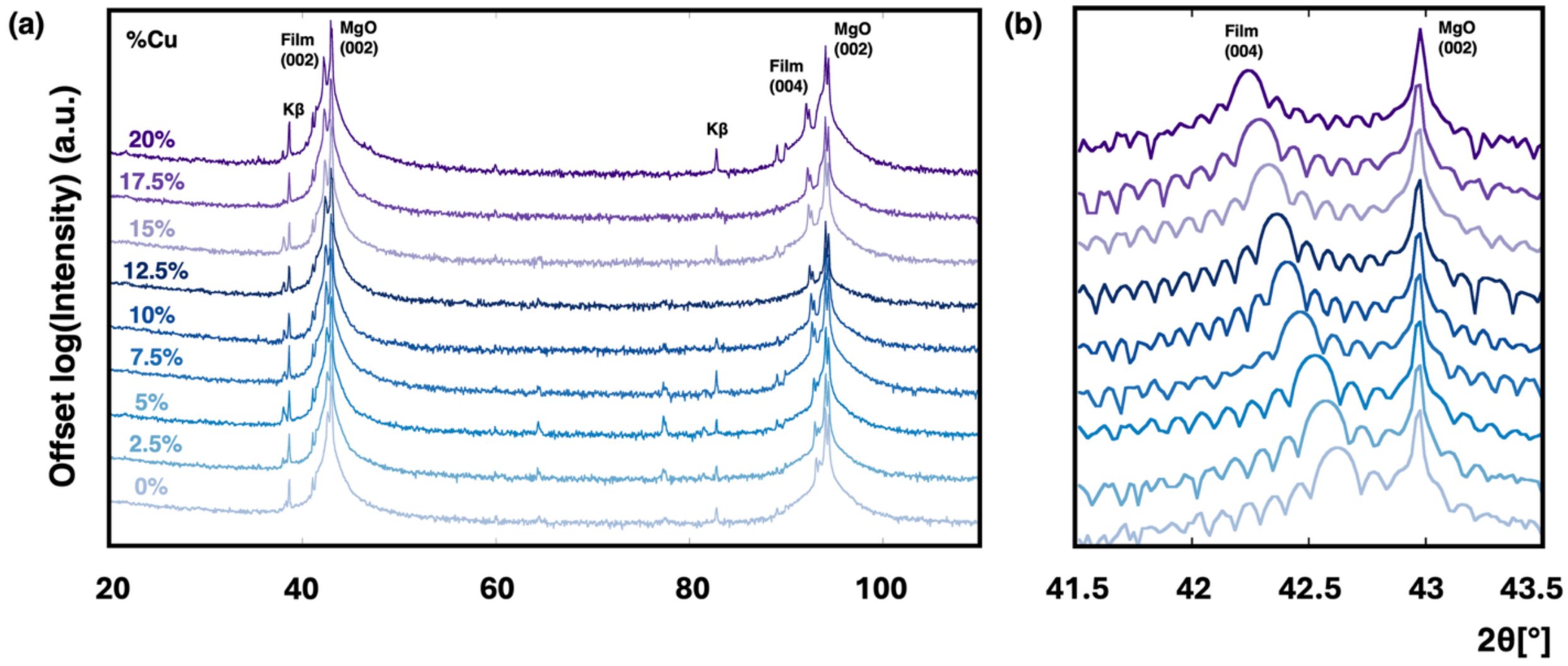

**Figure 2: Cu-driven lattice expansion in epitaxial $Cu_x(MgCoNiZn)_{1-x}O$ thin films.** (a) Bragg–Brentano high-definition X-ray diffraction (BBHD XRD) patterns for films grown on MgO at 500 °C with Cu concentrations ranging from 0% to 20%. (b) High-resolution parallel-beam XRD scans collected with a monochromator around the film and MgO (002) reflections, highlighting the Cu-induced peak shift and corresponding increase in the out-of-plane lattice parameter. Pronounced Pendellösung fringes indicate high crystalline quality across the series.

We next lower the substrate temperature to 300 °C while keeping all other growth conditions identical. Figure 3(a) shows θ-2θ BBHD XRD scans for the Cu-concentration series, arranged as in Figure 2(a), from 0% Cu at the bottom to 20% Cu, or equimolar J14, at the top in 2.5% increments. Based on the 500 °C series, we might expect the out-of-plane lattice parameter to increase with Cu concentration. Instead, the film peak shift toward higher 2θ with %Cu indicating out-of-plane lattice contraction – note also that the 0% Cu out of plane lattice spacing is similar for both temperatures. Similar to the 500 °C series, all films are pseudomorphic to MgO, unlike the 500 °C series, however, the films begin as tetragonal with c>a (here we refer to the out of plane direction as c), they end as tetragonal but with c<a, and exhibits a cubic "transition" structure in the vicinity of 10% Cu. At 500 °C the films start tetragonal and experience a monotonically increasing distortion. Figure 3(b) highlights this crossover through high-resolution scans around the MgO (002) reflection, it shows that all films retain high crystalline quality, and it shows that the out of plane lattice constant trend is strongest near the cross-over composition.

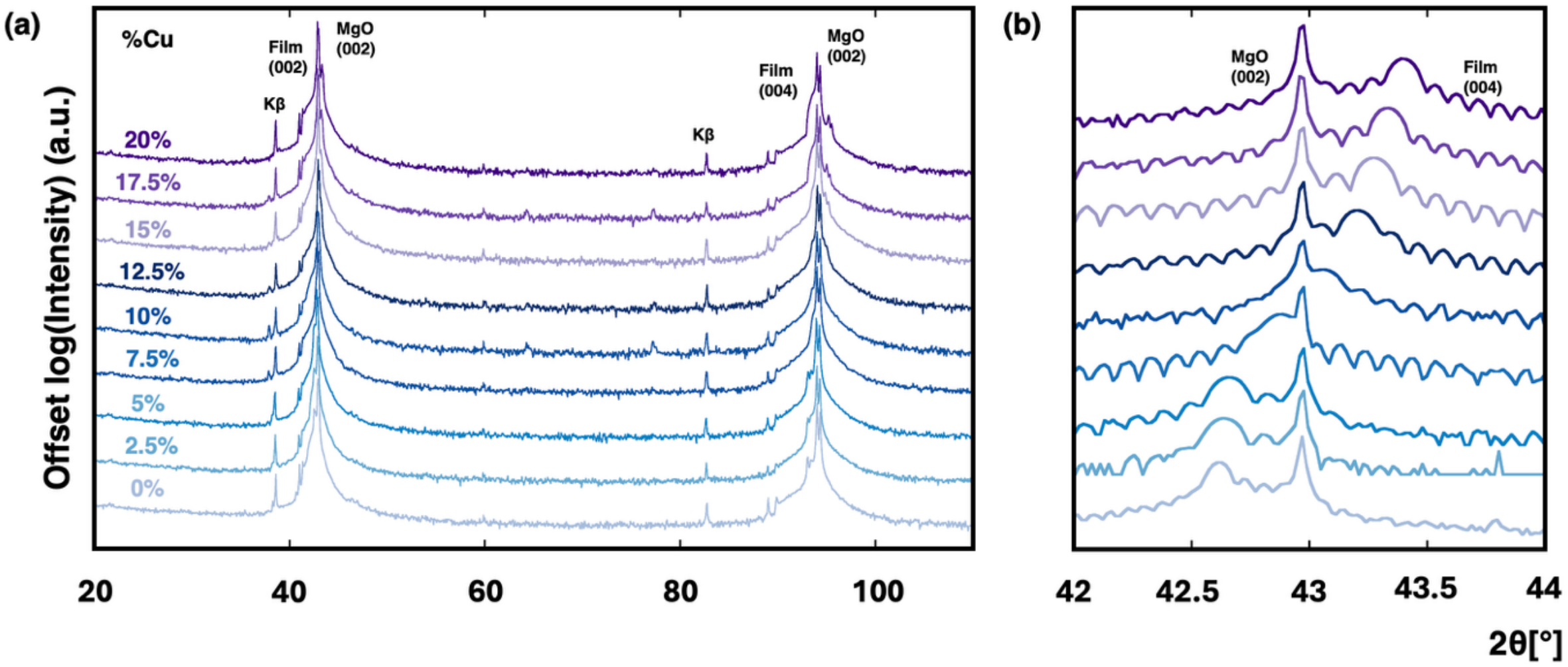

**Figure 3. Cu-driven lattice contraction in epitaxial $Cu_x(MgCoNiZn)_{1-x}O$ thin films.** (a) Bragg-Brentano high-definition X-ray diffraction (BBHD XRD) patterns for films grown on MgO at 300 °C with Cu concentrations ranging from 0% to 20%. (b) High-resolution parallel-beam XRD scans collected with a monochromator around the film and MgO (002) reflections, highlighting the Cu-induced peak shift and corresponding decrease in the out-of-plane lattice parameter. Pronounced Pendellösung fringes indicate high crystalline quality across the series.

Figure 4(a) consolidates the Cu-dependent out-of-plane lattice parameters and reveals contrasting behavior at 300 and 500 °C. The blue dotted line marks the MgO lattice parameter. At 500 °C, the lattice parameter increases nearly linearly by approximately 0.11% with each 2.5% Cu increment. At 300 °C, it decreases sharply and nonlinearly above 5% Cu and crosses below the MgO value near 10% Cu. The two series rapidly diverge ~5% Cu increases as quantified in Figure 4(b) with maximum strain ~2.6% for equimolar J14 for these growth conditions. This result suggests that Cu amplifies the temperature-dependent structural response and points to strong coupling between Cu-driven Jahn-Teller distortions and Co valence.

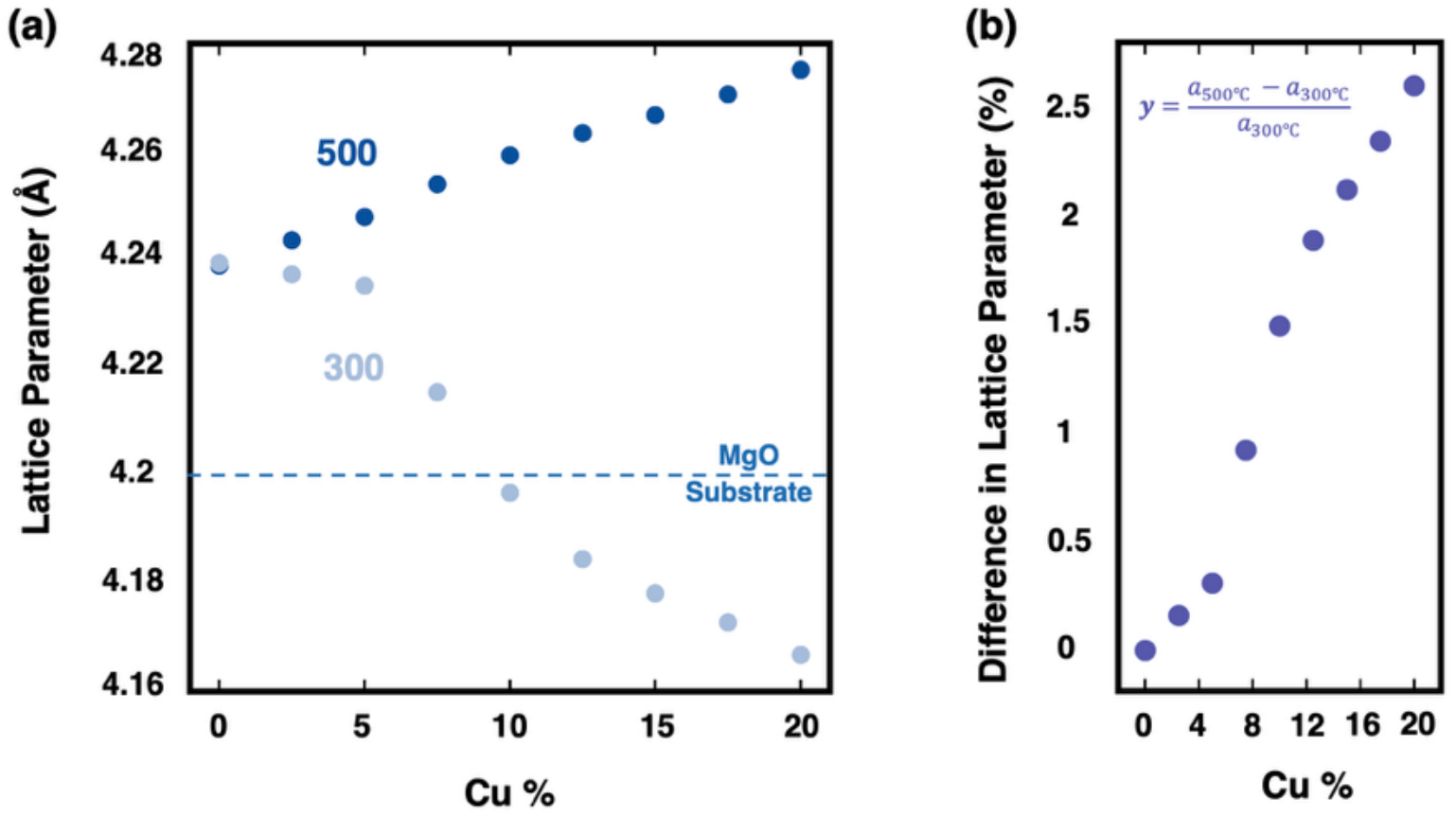

**Figure 4: Contrasting Cu-dependent lattice responses at 300 and 500 °C.** (a) Out-of-plane lattice parameter as a function of Cu concentration for films grown at 300 and 500 °C. The blue dotted line marks the MgO lattice parameter. (b) Relative difference between the out-of-plane lattice parameters of films grown at 500 and 300 °C.

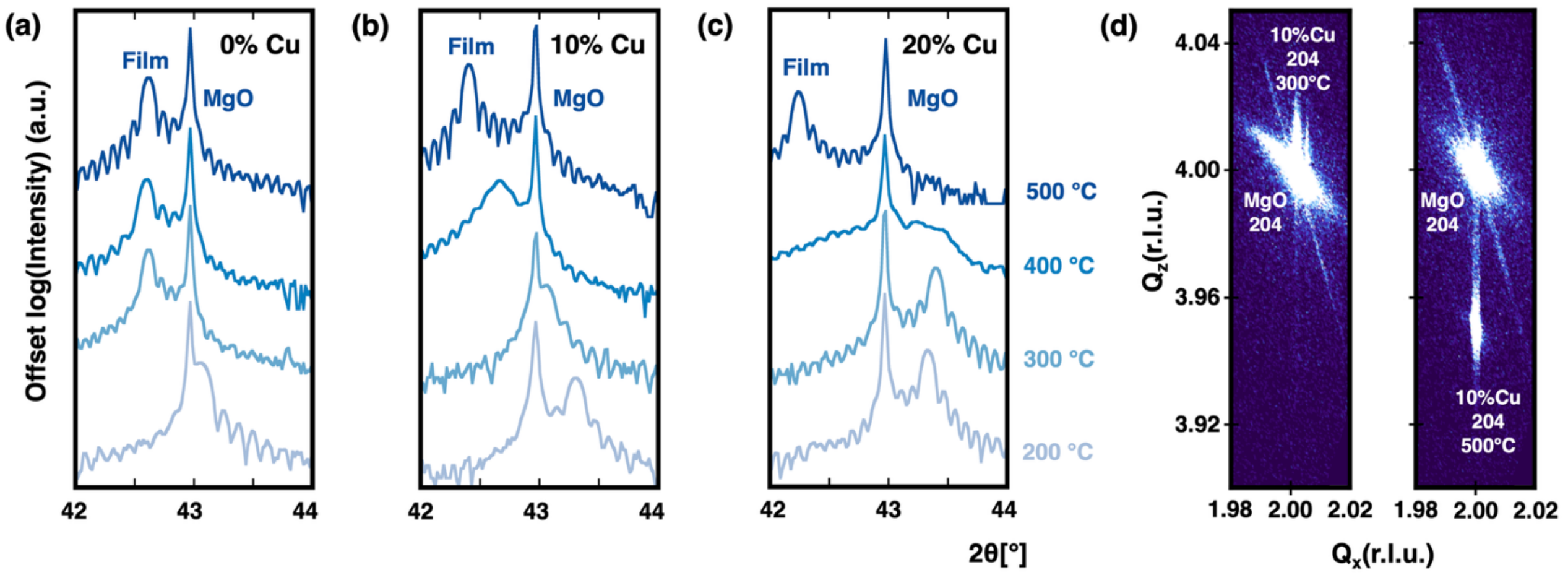

**Figure 5: Influence of Cu content and growth temperature on film structure.** High-resolution X-ray diffraction patterns of $Cu_x(MgCoNiZn)_{1-x}O$ films with (a) x = 0, (b) x = 0.10, and (c) x = 0.20 grown at temperatures ranging from 200 to 500 °C. (d) Reciprocal-space maps of the x = 0.10 films grown at 300 and 500 °C, confirming that both films remain in-plane lattice-matched to the MgO substrates and grow pseudomorphically.

To better illustrate the temperature dependent trends Figure 5(a-c) compares high-resolution θ–2θ XRD scans for films containing 0%, 10%, and 20% Cu grown from 200 °C to 500 °C in 100 °C increments. Without Cu, the film peak changes little between 300 °C and 500 °C and shifts appreciably only at 200 °C [Figure 5(a)]. At 10% Cu, the film peak progressively shifts toward lower 2θ as temperature increases, indicating out-of-plane lattice expansion [Figure 5(b)]. At 20% Cu, the lattice remains nearly constant from 200 to 400 °C before expanding abruptly between 400 and 500 °C, showcasing the largest temperature-driven shift in the series [Figure 5(c)]. We note that with increasing Cu, the “transition” state (where the films are nearly cubic) exhibits broader diffraction events, but re-sharpens at higher temperature. This suggests a possible mechanism of crystallographic disorder as the distorted CuO6 octahedra re-orient from a net in-plane (and likely in-plane random) to a net out-of-plane orientation as Co ions reduce from 3+ to 2+, and this transitional disorder becomes larger as Cu “unlocks” more net strain from the Co expansion thus participating in the transition

Figure 5(d) confirms that this response occurs without in-plane relaxation. Previous work establishes pseudomorphic growth for the 0% and 20% Cu compositions; here, we extend this result to 10% Cu. Reciprocal space maps of the 10% Cu films grown at 300 and 500°C show that the film reflection aligns with MgO along $Q_x$ at both temperatures. The 300°C film shows slight out-of-plane contraction relative to MgO, whereas the 650 °C film expands strongly along the growth direction. Despite this large out-of-plane expansion, the 650 °C film remains coherent in-plane, further exemplifying anomalous crystallinity.

## Conclusion and Outro

Collectively, these results reveal strong temperature-dependent coupling between Cu-driven Jahn-Teller distortions and Co valence. At 500 °C, Co remains predominantly $Co^{2+}$, allowing Jahn-Teller elongation to drive the out-of-plane lattice expansion. At 300 °C, we propose that increasing Cu strengthens local Jahn-Teller distortions and promotes further oxidation of $Co^{2+}$ to $Co^{3+}$. The resulting $Co^{3+}$-driven contraction outweighs the Cu-driven out-of-plane elongation and reverses the expected trend. The $CuO_6$ octahedra may also reorient locally, rotating their Jahn-Teller elongation axes toward the film plane and further reducing the out-of-plane lattice parameter. Together, Co oxidation and local $CuO_6$ reorientation provide a plausible origin for the nonlinear contraction at 300 °C and the 2.6% difference between the two temperature regimes for equimolar J14. Thus, the prevailing Co-centered picture is incomplete: Co valence alone cannot account for the J14 lattice response. Cu must be treated as a contemporary important structural control parameter because its Jahn-Teller distortions couple with Co valence to determine both the direction and magnitude of the lattice change.

## Methods

*Bulk Synthesis*

For bulk ceramic targets, we combine MgO (Sigma-Aldrich, 342793), CoO (Sigma-Aldrich, 343153), NiO (Sigma-Aldrich, 203882), CuO (Alfa Aesar, 44663), ZnO (Sigma-Aldrich, 96479). We mix and mill the powders with 5 mm yttrium stabilized zirconia media for 2.5 hours. We then press the powder into 2.5 cm diameter pellets at 60-100 MPa for 60 seconds. We react and sinter the ceramic targets in a muffle furnace at 950°C for 18 hours. We subsequently quench all samples from ~700 °C in air to prevent low-temperature phase segregation. We verify structural composition via X-ray diffraction (Panalytical Empyrean) using θ-2θ Bragg-Brentano HD scans with a PIXcel3D detector and identify phases with PANalytical HighScore. We monitor chemical compositions before and after sintering by X-ray fluorescence (Panalytical Epsilon 1).

*Thin Film Synthesis*

We bond MgO(100) substrates to the sample holder using silver paint and place the holder on an external hot plate for 20 min to dry the paint. We then transfer the holder and the mounted substrates to the chamber heater and preanneal the substrate under vacuum at 850 °C for 20 min. Next, we set the heater to the desired growth temperature and allow the substrates to equilibrate for an additional 20 min. During growth, we introduce $O_2$ at a flow rate of 50 sccm and maintain the chamber pressure at 50 mTorr. We operate the laser at 10 Hz for a total of 2,000 pulses while maintaining a target-to-substrate distance of 5 cm and a fluence of 2 $Jcm^{-2}$. Immediately after growth, we rapidly cool the samples by transferring them to room temperature load lock to set there for 10 minutes before removing them from the chamber to cool in air. In addition to BBHD X-ray diffraction scans, we collect high-resolution scans using a 2xGe hybrid monochromator on the incident-beam side, an additional 2xGe crystal analyzer on the diffracted-beam side with a proportional detector.

**Acknowledgments**
The authors gratefully acknowledge support from NSF MRSEC DMR-2011839. The authors would also acknowledge the help of Matthew Hammell in synthesizing the bulk ceramics.

**Conflict of Interest**
The authors declare no conflicts of interest.

**Data Availability**
The data supporting the findings of this study are available from the corresponding author upon reasonable request.